\documentclass[preprint,aps]{revtex4}

\usepackage{graphicx}
\usepackage{dcolumn}
\usepackage{bm}

\begin{document}
\title{Generating Cosmological Solutions from Known Solutions}
\altaffiliation{Published in: ``Mathematical and Quantum Aspects of
Relativity and Cosmology'', S. Cotsakis and G. W. Gibbons (Eds.)
(Springer Verlag, Berlin, 2000) pp. 191-213}
%
%
%
%
\author{Hernando Quevedo and Michael P. Ryan, Jr.}\email{quevedo, ryan@nuclecu.unam.mx}
%
\affiliation{Instituto de Ciencias Nucleares, UNAM, A. Postal
70-543, \\
M\'exico 04510 D.F., Mexico}

\begin{abstract}
We consider three methods by which one can generate new cosmological
models.  Two of these are based on the Lorentzian structure of spacetime.
In a Lorentzian manifold there can exist horizons that separate regions of
spacetime that can be interpreted as cosmological models from others that
have the character of ``black holes.''  A number of well known solutions 
of this type can be used to generate both known cosmological models and
others that do not seem to have been recognized.  Another method based on
the Lorentzian character of spacetime is to simply interchange some space
variable with time and try to restructure the metric to make a viable
cosmology.

A more broad-ranging method is the use of modern solution-generating 
techniques to construct new models.  This method has been widely
used to generate black hole solutions, but seems not to have been so
widely used in cosmology.  We will discuss examples of all three methods.
\end{abstract}

\maketitle

\section {Introduction}
The standard model of cosmology assumes that the universe today is
homogeneous and isotropic, which means that there are six Killing vectors
of the manifold that give the isometries that realize these symmetries.
However, in earlier stages of the universe it is assumed that there might
have been large amounts of anisotropy and inhomogeneity that, by some
physical process, could have been reduced to the point where today we
observe the standard model (for a compendium of reasons for considering
inhomogeneous models, see Ref. \cite{kras}).
Anisotropic and inhomogeneous metrics have
fewer and fewer Killing vectors, and we could, in principle, arrive at
a completely general solution of the Einstein equations as a
cosmological model.  At some point it will become difficult to distinguish
between a cosmological model and any other general solution of the
Einstein equations, except as a matter of interpretation.  In fact, even
an inhomogeneous cosmology with two spacelike Killing vectors is
difficult to distinguish (locally) from
gravitational waves propagating in one space direction.  One method
that has been used to separate cosmological models from pure wave solutions
is the prescription of their global topology.  If we insist that cosmologies
have compact $t =$ constant surfaces, then a number of gravitational wave
solutions can be made into cosmologies by compactifying in certain directions
of the system.  We will give examples of this procedure below.  If we allow 
open universes, whether we call a solution a cosmological model or not
is again a matter of interpretation.

With these caveats we can give a number of methods by means
of which one can transform known solutions of the Einstein equations
into cosmological models.  These methods can be broken down into three
broad classes:
\vskip 10 pt

\hskip 1 true cm 1) Horizon methods

\hskip 1 true cm 2) Causal structure methods

\hskip 1 true cm 3) Mapping methods
\vskip 10 pt

The first of these methods has a long history, even though it never seems
to have been thought of as a ``method.''  Perhaps the oldest example is the 
deSitter metric \cite{deS}, although
it is a somewhat degenerate example.  Originally
deSitter proposed a solution to the Einstein equations with cosmological
constant $\Lambda$ of the form \cite{redeS}

\begin{equation}
ds^2 = -\left ( 1 - {{\Lambda r^2}\over {3}}\right )dt^2 + {{dr^2}\over 
{\left (1 -
{{\Lambda r^2}\over {3}}\right )}} + r^2 (d\theta^2 + \sin^2 \theta 
d\varphi^2). \label {des}
\end{equation}
This seems to be a static metric similar in form to the Schwarzschild
metric, and even has a ``singularity'' (which caused much comment at the
time) at $r = \sqrt{3/\Lambda}$.  Of course, this singularity is just a
horizon where the light cones tip over sufficiently that for $r > 
\sqrt{3/\Lambda}$, $\partial/\partial r$ becomes timelike and $\partial/
\partial t$ spacelike.  For large $r$, then, we
can rename the coordinates ($r
\rightarrow \tilde t$, $t \rightarrow \tilde r$), and the new metric is 
\begin{equation}
ds^2 = -{{d\tilde t^2}\over {{{\Lambda \tilde t^2}\over {3}} - 1}} + 
\left ( {{\Lambda
\tilde t^2}\over {3}} - 1 \right ) d\tilde r^2 + \tilde t^2 (d\theta^2 +
\sin^2 \theta d\varphi^2), \label {des1}
\end{equation}
which is an obvious cosmological model.
It was ``recognized'' (the reason for using the quotation marks will
become obvious below) relatively quickly that this
was indeed a cosmological model by
making use of the coordinate transformation 
\begin{eqnarray}
\tilde t = r e^{\sqrt{\Lambda/3}T}, \quad \tilde r = T -
{{1}\over {2}}\sqrt{{{3}\over {\Lambda}}} \ln
\left ( {{\Lambda r^2 e^{2\sqrt{\Lambda/3}T}}\over {3}} - 1 \right )\ ,
\nonumber \\ 
r\sin \theta \cos \varphi = X, \qquad
r\sin \theta \sin \varphi = Y, \qquad
r\cos \theta = Z \ , \label{cot}
\end{eqnarray}
which makes the metric
\begin{equation}
ds^2 = -dT^2 + e^{2\sqrt{\Lambda/3}T}[dX^2 + dY^2 + dZ^2], \label {tol}
\end{equation}
which is a $k = 0$ Friedmann-Robertson-Walker metric.

\begin{figure}
\begin{center}
\includegraphics[width=.7\textwidth]{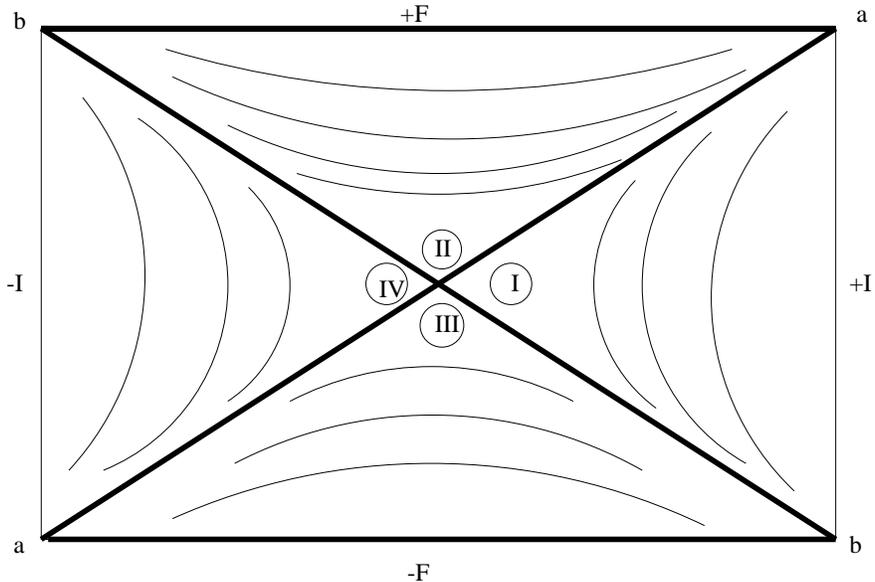}
\end{center}
\caption [] {A generic Penrose diagram which will be used for several
metrics of the article}
\label{pdia}
\end{figure}

Here we should study the Penrose diagram of this metric.  In Figure
\ref{pdia}
we give a Penrose diagram that will cover several cases that we will
discuss later.  Because of this we will give only generic values in the
figure, and each case will correspond to different values of these parameters.
In the deSitter case, the lines $aa$ and $bb$ correspond
to $r = \sqrt{3/\Lambda}$, and both $\pm I$ to $r = 0$.  In region I the
spacelike Killing vector $\partial /\partial r$ generates spacelike
surfaces that are the $t =$ constant surfaces of (\ref{des}).  In region 
II the same Killing vector is now $\partial /\partial \tilde t$ (timelike)
and generates timelike hypersurfaces $\tilde r =$ constant.  The horizon
splits the region comprised by both regions I
and II in two.  Region I might be called a ``black hole''
region since the metric is reminiscent of the Schwarzschild metric, and
has the spacelike Killing vector $\partial /\partial r$, while region II
is a ``cosmological'' region with metric (\ref {tol}).  In the cosmological
region there seem to be  ``singularities'' at $T = \pm \infty$, but we can 
see that [note that $T = \tilde r + {{1}\over {2}}\sqrt{{{3}\over {\Lambda}}}
\ln \left (
{{\Lambda \tilde t^2}\over {3}} - 1 \right )$] the singularity at $T =
-\infty$ is just the horizon at $\tilde t = \sqrt{3/\Lambda}$,
so it is nothing more than a null surface separating two parts of the 
spacetime.

The problem here is that the deSitter metric has such a high degree of 
symmetry (a space of constant curvature with 10 Killing vectors)
that the distinction between regions I and II is artificial.
In fact, the coordinate transformation (\ref{cot}) (with $\tilde t 
\rightarrow r$ and $\tilde r \rightarrow t$ and ${{\Lambda \tilde t^2}\over
{3}} - 1 \rightarrow {1 - {\Lambda r^2}\over {3}}$) is equally valid
in region I, and, in fact, this was what
was recognized by Lemaitre \cite{lem} and Robertson \cite{rob}
in transforming the deSitter metric into the form that was
eventually used in the steady-state model.
Actually, even Minkowski space (also a space of constant curvature with 10
Killing vectors) can be looked at in the same way, as we will see
in Sec. 2.

In Sec. 2 we will consider several metrics where horizon methods can be used
to construct cosmological models that do not have the disadvantages of the
deSitter metric, that is, there exists no coordinate transformation that
is valid for the static region which transforms it into a cosmological model.
The fact that ``singularities'' in the new cosmological models may just
be null surfaces where one passes from one coordinate patch to another, will,
of course, still be a feature, although true curvature singularities may
also exist.  One point that should be mentioned is that most of
the models generated by this method that we will consider are vacuum models.
There are some electrovac models that can be thought of as being
generated by horizon methods that we will mention, but it seems not to be
known which matter-filled cosmological models can be generated by this method.

The second class of methods, causal structure methods,
is similar to the previous one, but there is
no horizon to distinguish different regions of spacetime.  The paradigm
of this method is the unpolarized Gowdy $T^3$ vacuum
spacetime \cite{gow} which will be discussed in
more detail below.  The idea behind this method is to begin with a known 
solution and simply rename some spacelike variable as $t$, and the timelike
variable as a spacelike variable (we can call this $t \leftrightarrow r$),
and then change the causal structure of the resulting metric so that the
a vector tangent to the ``time'' direction is truly timelike, and one 
tangent to the new ``space'' direction is spacelike.  There are a number
of problems with this approach.  The first is that determining the 
causal structure change is not necessarily trivial.  In a diagonal metric
this difficulty can usually be handled by simply changing the signs of
two of the metric components, but in a more general metric the procedure
may be more complicated.  The most important difficulty, however, is that
the new metric is not guaranteed to be a solution to the Einstein equations.
For matter filled models, where physical quantities can be used to define
spacelike and timelike surfaces, the new metric will almost certainly not
be a solution.  For vacuum metrics it is difficult to tell.
As with horizon methods, the known
solutions of this type are vacuum wave solutions, and they have rather simple 
structures, and give inhomogeneous cosmological models.  Here, of course, one
has the problem of distinguishing between gravitational wave solutions and
inhomogeneous models.

A very simple example of this procedure is the plane wave \cite{pla}
\begin{equation}
ds^2 = L^2(u)(e^{2\beta(u)}dx^2 + e^{-2\beta(u)}dy^2) + dz^2 - dt^2,
\label {pla}
\end{equation}
where $u = t - z$.  If we make the coordinate change $t \rightarrow \tilde 
z$, $z \rightarrow \tilde t$ and change the signs of $dz^2$ and $dt^2$, we
find ($\tilde u = \tilde t - \tilde z$)
\begin{equation}
ds^2 = L^2(-\tilde u)[e^{2\beta (-\tilde u)}dx^2 + e^{-2\beta(-\tilde u)}dy^2]
- d\tilde t^2 + d\tilde z^2. \label {newpl}
\end{equation}
Since $L$ and $\beta$ are arbitrary function of their arguments, we can 
remove the minus signs in the arguments of
$L$ and $\beta$ in (\ref{newpl}) and obtain
(\ref{pla}) again.  Because of the high symmetry of the problem, this 
procedure gives the same metric and Einstein equations, and there is no
question that any solution of the Einstein equations for (\ref{pla}) gives
a solution to (\ref{newpl}).  Here, of course, one runs into the problem
of defining a cosmological model.  Is either of (\ref{pla}) or (\ref{newpl})
a gravitational wave or an inhomogeneous cosmological model?  Either one
is both, depending on how we interpret them.  If, however, we insist that
a cosmological model have compact $t =$ const. sections, then we can 
compactify (\ref{newpl}) in the space directions by just
making $0 \leq x, y, z \leq 2\pi$ and
identifying the points at $0$ and $2\pi$, giving the manifold a $T^3$
topology.  This changes the boundary conditions on the functions $L$ and
$\beta$ in that one must have $L(t, 0) = L(t, 2\pi)$ and $\beta (t, 0) =
\beta (t, 2\pi)$.  Since the only Einstein equation is 
\begin{equation}
L^{\prime \prime}(\tilde u) + [\beta^{\prime} (\tilde u)^2] L = 0,
\label {eeq1}
\end{equation}
$^{\prime} = d/d\tilde u$, $\beta$ is arbitrary except for the boundary
conditions, and $L$ obeys a simple equation that has solutions that
satisfy the boundary conditions.  Here one must be careful not to change
coordinates back to the original metric, since we will generate a spacetime
with closed timelike lines, a problem with cosmological models
generated in this manner.  We will discuss examples and their topology
in Sec. 2.

The third method consists of taking known solutions and mapping them to
new solutions.  This technique is similar in spirit to conformal mapping.
Any analytic function of one complex variable,
$w = f(z)$ maps $z = x + iy$ to $w = u(x, y) +
iv(x, y)$, where, if $\psi (x, y)$ is a solution to Laplace's equation,
$\Psi [u(x, y), v(x, y)]$ is also a solution.  This idea exploits the
invariance of Laplace's equation in two dimensions
under the two-dimensional conformal group. If one can find groups
under which the Einstein equations are invariant (or
under which some class of Einstein equations for special systems are
invariant), one can generate new solutions from old.  This technique is 
of wide applicability, and has been exploited heavily in general relativity
to find new exact solutions.  It is by no means specific to cosmological
models, but it can be applied in cosmological situations, and has,
perhaps, been underutilized in this field and deserves
more attention there.

In the specific case of stationary axisymmetric spacetimes, early 
generating methods such as the Kerr-Schild {\it ansatz}
\cite{kerr1,newman1,kinn1,pledem1}, 
the complex transformation method \cite{newjan,newman2,demnew}, 
and Hamilton-Jacobi separability \cite{carter1} were used to
derive new solutions, but all of them are contained in the 
charged Kerr-Taub-NUT \cite{ktn} class (with cosmological constant).
An important development for the derivation of exact stationary axisymmetric 
solutions was made by Ernst \cite{ernst1,ernst2}, who obtained 
a new representation of the corresponding field equations which 
is independent of the coordinates chosen and, therefore, allows 
one to investigate the symmetries involved and find new solutions.
Ernst also proposed two generating techniques that were later 
enlarged and unified by Kinnersley \cite{kinn2,kinn3}.
Using the compact Ernst formulation of the field equations it
was also possible to obtain new solutions
\cite{tomsat1,tomsat2,yamhori,yam1,yam2} possessing a certain 
prescribed polynomial form of the Ernst potential. Modern 
solution generating techniques involve Lie groups of 
transformations or B\"acklund transformations. 
The first such group was found by Geroch \cite{ger1,ger2} and 
generalized by Cosgrove \cite{cos1,cos2}. The Geroch group
was investigated very intensively and it was found that it 
possesses subgroups that preserve asymptotic 
flatness \cite{kinn2,kinchi1,kinchi2,kinchi3,kinchi4}. 
Today, there exist three main solution generating tecniques: HKX 
(Hoenselaers, Kinnersley, Xanthopoulos) transformations \cite{hkx1,hkx2},
which are based upon special
subgroups of the Geroch group, B\"acklund transformations 
\cite{harr1,neug1}, which are applied directly to the Ernst 
potential, and the inverse scattering method 
\cite{belzak1,belzak2}, which is based on a reformulation of
the nonlinear field equations as a linear eigenvalue problem. 
Later on it was shown \cite{cos3,cos4} 
that all these techniques are related to one another, and can be 
used to generate the same type of solutions (for an introductory
review of the solution generating techniques and the 
relationships between them see \cite{que1}). 

The mapping method is by no means specific to stationary 
axisymmetric metrics, but it can be applied to any spacetime 
characterized by two or more commuting Killing vector fields as in the
case of the cosmological models investigated in this work. 
We will give some examples in Sec. 3 and a guide to possible new
uses of the technique.

The paper is organized as follows.  Section 2 will give examples of horizon
and causal structure methods and solutions obtained using them.  Section 3
will give one example of mapping methods and discuss others.
Finally, Section 4 will discuss possible new directions in the use of known
solutions to find new cosmologies.

\section{Horizon and Causal Structure Methods}
\subsection{Horizon Methods}
There are a large number of solutions of the Einstein equations that have
cosmological coordinate patches as well as ``black hole'' regions that
are separated by horizons.  The major problem with the cosmologies generated
in this way is they are incomplete manifolds, and what have in the past been
interpreted in some of them as singularities are just the horizons that
separate one part of the complete manifold from the other.  A second
problem, as we have seen, is that in certain cases the complete manifold
is of such high symmetry that is is impossible to distinguish between
the cosmological region and the ``black hole'' region.  The most blatant
example of the second problem is just ordinary Minkowski space.

If we return to Figure \ref{pdia}, we can think of this diagram as
a picture of Minkowski space with the meeting point of lines $aa$ and
$bb$ an arbitrarily chosen origin $O$. The usual metric is
\begin{equation}
ds^2 = -dT^2 + dx^2 + dy^2 + dz^2.
\label{minky}
\end{equation}
In region I the hyperbolic lines are
just the usual orbits of a particle with constant acceleration that form the
basis of Rindler \cite{rin} space, and
if the lines are given by $x = \sqrt{2r - 1}\cosh t$ and $T =
\sqrt{2r - 1}\sinh t$, $z = y =$ const., the metric becomes
\begin{equation}
ds^2 = -(2r -1)dt^2 + {{dr^2}\over {2r - 1}} + dy^2 + dz^2.
\label{rind}
\end{equation}
In this coordinate system the metric is very similar in form to the
Schwarzschild metic, and there is a coordinate singularity at $r = 1/2$,
which is a horizon similar to that of the deSitter metric.

In Figure \ref{pdia} the line $bb$ is the surface $r = 1/2$, and for
$r < 1/2$ we can make the $t \leftrightarrow r$ coordinate change and the
metric becomes
\begin{equation}
ds^2 = -{{1}\over {1 - 2t}}dt^2 + (1 - 2t)dr^2 + dy^2 + dz^2,
\label{rinmil}
\end{equation}
a cosmological model.  If we change coordinates using $t = {{1}\over {2}}
[1 - \tau^2\{\cosh^2 \rho - \sinh^2 \rho \sin^2 \theta \cos^2 \varphi \}]$,
$r = \tanh^{-1} [\tanh \rho \sin \theta \cos \varphi]$, $y = \tau \sinh \rho
\sin \theta \sin \varphi$, $z = \tau \sinh \rho \cos \theta$,
the metric (\ref{rinmil}) becomes the Milne universe, a flat cosmological
model with metric
\begin{equation}
ds^2 = -d\tau^2 + \tau^2[d\rho^2 + \sinh^2 \rho(d\theta^2 +
\sin^2 \theta d\varphi^2)],
\label{miln}
\end{equation}
a $k = -1$ FRW metric.  Of course, the manifold is nothing more than flat
space and the breaking up of the manifold into regions I-IV is an
observer-dependent phenomenon due to singling out the origin $O$ as the
position of a special observer, while there is actually no physical
reason that this point is priveleged over any other point of the manifold,
and there is no real difference between any of the four regions.  Basically,
we can say that the same is true of deSitter space, which is just a constant
curvature space everywhere, and the ``cosmological'' region has nothing to
distinguish it from any other region of the manifold.

There are a number of metrics which are not as degenerate as the Minkow- ski
and deSitter cases whose cosmological regions are physically
distinguishable from the ``black hole'' regions, and the resulting
cosmological models have long been known and are named.  Perhaps the simplest
of these is the Kantowski-Sachs-Schwarzschild manifold.  Here we can still
use the generic Figure \ref{pdia} to represent the
Penrose diagram of this metric,
with region I representing the Schwarzschild coordinate patch where the
hyperbolic lines are $r =$ const. lines and the lines $bb$ is the horizon
at $r = 2m$.  In region II the hyperbolic lines are still $r =$ const. lines,
but they no longer represent spacelike surfaces.  For $r > 2m$ the metric
can be written as
\begin{equation}
ds^2 = -\left ( 1 - {{2m}\over {r}}\right )dt^2 + {{1}\over
{1 - {{2m}\over {r}}}}dr^2 + r^2 (d\theta^2 + \sin^2 \theta d\varphi^2),
\label{schw}
\end{equation}
but for $r < 2m$ the Killing vector $\partial /\partial r$ is no longer
spacelike, and we may make the transformation $r \leftrightarrow t$ mentioned
in Sec. 1, and we find
\begin{equation}
ds^2 = -{{1}\over {{{2m}\over {t}} - 1}} dt^2 + \left ( {{2m}\over {t}} - 1
\right ) dr^2 + t^2 (d\theta^2 + \sin^2 \theta d\varphi^2),
\label{ks}
\end{equation}
which is an obvious cosmological model, the vacuum Kantowski-Sachs model.
The fact that at $r = 0$ ($t = 0$ in the new coordinate system) which is
represented by the line at $+F$ is a true curvature singularity distinguishes
region II from region I and makes it impossible to find a global coordinate
transformation that makes this cosmological model indistinguishable from
the ``black hole'' region.  Of course, however, this cosmological model,
and all cosmological models generated by horizon methods, have the problem
that what has been regarded as singularities in these models may not be
a curvature singularity, but only a horizon where we pass from one coordinate
patch to another.  In the case of the Kantowski-Sachs solution the $t = 0$
singularity is one of the usual singularities of the model, and it is a
true curvature singularity.  The other ``singularity'' at $t = 2m$ is only
a null surface.

While the Kantowski-Sachs solution, even though it was originally discovered
by means of studies of groups of motion that were not transitive on three
surfaces, and constituted a generalization of the Bianchi cosmological
models, and was immediately recognized as the Schwarzschild solution
inside the horizon, in the next model we will discuss the two regions
on either side of the horizon were studied separately.  The ``black
hole'' region was the NUT space of Newman, Tamburino and Unti \cite{newman1},
and the cosmological region was the Taub cosmology \cite{taub}.
Misner \cite{chas} showed that these two solutions could be seen to
be part of a larger manifold (up to topological questions which will
be discussed below).  In the black hole region one can write the NUT
metric as \cite{kramet}
\begin{eqnarray}
ds^2 = -{{r^2 - 2mr -l^2}\over {r^2 + l^2}}(dt + 2l\cos \theta d\varphi)^2
+ {{r^2 + l^2}\over {r^2 - 2mr - l^2}}dr^2 \ \nonumber \\
\qquad \qquad + (r^2 + l^2)(d\theta^2 +
\sin^2 \theta d\varphi^2).
\label{nut}
\end{eqnarray}
For the region of this metric where $0 < r < m(1 + \sqrt{1 + l^2/m^2})$,
the $r \leftrightarrow t$ transformation gives
\begin{eqnarray}
ds^2 = {{l^2 + 2mt - t^2}\over {t^2 + l^2}}(dr + 2l \cos \theta d\varphi)^2
- {{t^2 + l^2}\over {l^2 + 2mt - t^2}}dt^2 \ \nonumber \\
\qquad \qquad + (t^2 + l^2)(d\theta^2 + \sin^2
\theta d\varphi^2),
\label{taub}
\end{eqnarray}
which is an anisotropic cosmological model.  The major problem with this
model is that in the cosmological sector we can rewrite the metric on
$t =$ const. surfaces by making the coordinate transformation
\begin{eqnarray}
y & = & \theta - \pi/2 \ , \label{t1a} \\
z & = & r/2l \ , \label{t1b} \\
x & = & \varphi \ , \label{t1c}
\end{eqnarray}
which gives the three-dimensional line element as
\begin{equation}
d\sigma^2 = {{4l^2(l^2 + 2mt - t^2)}\over {t^2 + l^2}}(dz - \sin y dx)^2
+ (t^2 + l^2)(dy^2 + \cos^2 y dx^2),
\label{taub3}
\end{equation}
which can be written as
\begin{equation}
d\sigma^2 = {{4l^2(l^2 + 2mt - t^2)}\over {t^2 + l^2}}(\omega^3)^2 +
(t^2 + l^2)[(\omega^1)^2 + (\omega^2)^2],
\label{ntb3}
\end{equation}
where $\omega^3 = dz - \sin y dx$, $\omega^1 = \cos y \cos z dx - \sin z dy$,
$\omega^2 = \cos y \sin z dx + \cos z dy$.  These differential forms are the
invariant one-forms of the Bianchi type IX cosmological models, the invariant
one-forms on $S^3$.  This means that the ``natural'' topology of the
three-surfaces given by (\ref{ntb3}) is that of $S^3$, with $0 \leq z
\leq 4\pi$, $0 \leq x \leq 2\pi$, $0 \leq y \leq \pi$.
If one maintains this topology on passing through
the horizon, the vector $\partial/\partial z$ becomes timelike, and the $S^3$
topology implies the possibility of closed timelike lines.  This topological
difficulty is present in many cosmological models generated by horizon
methods, as we will see below.

The metrics we have discussed seem to be the only ones where the cosmological
models have been noticed explicitly to be a part of a larger manifold that
has non-cosmological sectors. For example, it is only recently that the
Kerr metric inside its horizon has been considered as a cosmological model.
Of course, the metric in this part of the Kerr manifold has been studied
\cite{ori}, but there seems to have been no attempt to identify it as a
cosmological model.  In Boyer-Lindquist coordinates the Kerr metric has
the form
\begin{eqnarray}
ds^2 = - {{r^2 - 2Mr + a^2}\over {r^2 + a^2\cos^2 \theta}}[dt - a\sin^2
\theta d\phi]^2 + {{\sin^2 \theta}\over {r^2 + a^2\cos^2 \theta}}[(r^2 +
a^2) d\phi - adt]^2 \ \nonumber \\
\qquad \qquad + {{r^2 + a^2\cos^2 \theta}\over {r^2 - 2Mr + a^2}}dr^2 +
(r^2 + a^2\cos^2\theta)d\theta^2
\label{kbl}
\end{eqnarray}
This metric has two horizons, $r_{\pm} = M \pm \sqrt{M^2 - a^2}$, and
beyond the outer horizon, $r_{+}$, $\partial/\partial r$ is a spacelike
vector, and the metric represents a spinning black hole.  At $r_{+}$
the light cones tip over to the point where $\partial /\partial r$ becomes
timelike and we can make the transformation $t \leftrightarrow r$.  
Unfortunately, at the inner horizon $r_{-}$ the light cones tip back to
the point where $\partial /\partial r$ becomes spacelike again.  In the
region $r_{-} < r < r_{+}$ the Kerr metric becomes a cosmological model,
\begin{eqnarray}
ds^2 = {{2Mt - t^2 - a^2}\over {t^2 + a^2 \cos^2 \theta}} [dr -
a \sin^2 \theta d\phi]^2 + {{\sin^2 \theta}\over {t^2 + a^2 \cos^2 \theta}}
[(t^2 + a^2)d\phi - adr]^2 \ \nonumber \\
\qquad \qquad -{{t^2 + a^2 \cos^2 \theta}\over {2Mt - t^2 - a^2}}dt^2 + (t^2 + 
a^2 \cos^2 \theta)d\theta^2.
\label{kgow}
\end{eqnarray}
This metric, with the simple transformation \cite{ory},
\begin{equation}
t = \alpha[\sqrt{1 - \beta^2} \cos (e^{-\tau}) + 1],
\label{kertr}
\end{equation}
$\alpha = M$, $\beta = a/M$ ($0 \leq \beta \leq 1$) transforms (\ref{kgow})
into
\begin{eqnarray}
ds^2 = e^{-\lambda/2} e^{\tau/2} (-e^{-2\tau}d\tau^2 + d\theta^2) 
+ \alpha \sqrt{1 - \beta^2}\sin(e^{-\tau})[e^P d\delta^2\ 
\nonumber \\
\qquad \quad +
2e^P Q d\delta d\phi +
(e^P Q^2 + e^{-P} \sin^2 \theta )d\phi^2],
\label{gow}
\end{eqnarray}
where
\begin{equation}
\lambda =  \tau - 2\ln (\alpha^2 \{[\sqrt{1 - \beta^2} \cos (e^{-\tau})
+ 1]^2 + \beta^2 \cos^2 \theta\} )\ ,
\label{kga}
\end{equation}
\begin{eqnarray}
P = \ln [(1 - \beta^2) \sin^2 (e^{-\tau}) + \beta^2 \sin^2 \theta ] -
\ln [\alpha \sqrt{1 - \beta^2} \sin (e^{-\tau})]\ \nonumber \\
\qquad \quad -\ln ([\sqrt{1 - \beta^2} \cos (e^{-\tau}) + 1]^2 + \beta^2
\cos^2 \theta),
\label{kgb}
\end{eqnarray}
\begin{equation}
Q = -{{2\alpha \beta \sin^2 \theta [\sqrt{1 - \beta^2} \cos (e^{-\tau}) +
1]}\over {(1 - \beta^2)\sin^2 (e^{-\tau}) + \beta^2 \sin^2 \theta}}\ .
\label{kgc}
\end{equation}
The metric (\ref {gow}) is a Gowdy cosmological model with $S^1 \times
S^2$ topology \cite{gow}. 

The solution given in Eqs. (\ref {kga} -\ref {kgc}) is a function of $\theta$
for $0 \leq \theta \leq \pi$ and is valid for values of $\tau$ that 
correspond to the part of the manifold between $r_{-}$ and $r_{+}$, that is,
for $-\ln (\pi) \leq \tau \leq +\infty$, (note that $t$ takes the values
$M \pm \sqrt{M^2 - a^2}$ at the two limiting values of $\tau$).

\begin{figure}[tb]
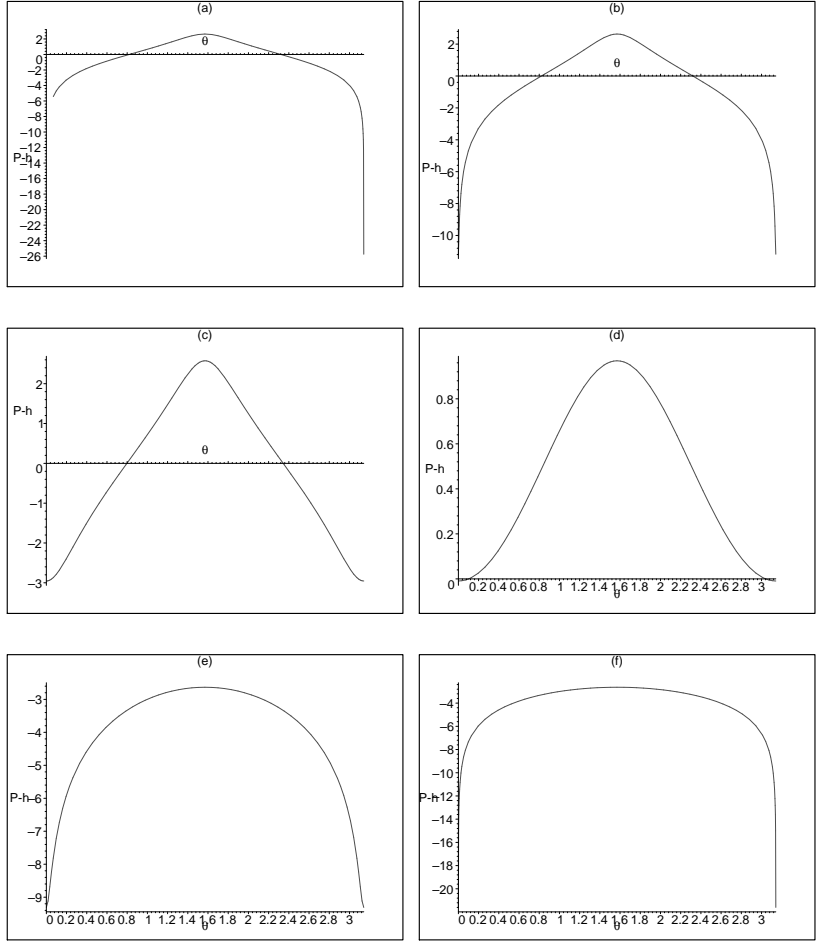

\[
\begin{array}{cc}
{\includegraphics[width = 1.50in, angle = 270]{gow7f.eps}} &
{\includegraphics[width = 1.50in, angle = 270]{gow6f.eps}}\\
{\includegraphics[width = 1.50in, angle = 270]{gow5f.eps}} &
{\includegraphics[width = 1.50in, angle = 270]{gow4f.eps}}\\
{\includegraphics[width = 1.50in, angle = 270]{gow1f.eps}} &
{\includegraphics[width = 1.50in, angle = 270]{gow3f.eps}} \cr
\end{array}
\] 
\caption[]{The evolution in $\tau$ of $P - h$ as a function of $\theta$
from Eqs. (\ref{kga}-\ref{kgc}) for $\alpha = 1$, $\beta = 1/2$.  Fig. (a)
corresponds to $\tau = -\ln (\pi)$, (b) to $\tau = -1.144$, (c) to $\tau =
-1.1$, (d) to $\tau = -0.75$, (e) to $\tau = +5$, (f) to $\tau = +10$}
\label{kgow1}
\end{figure}

\begin{figure}[tb]
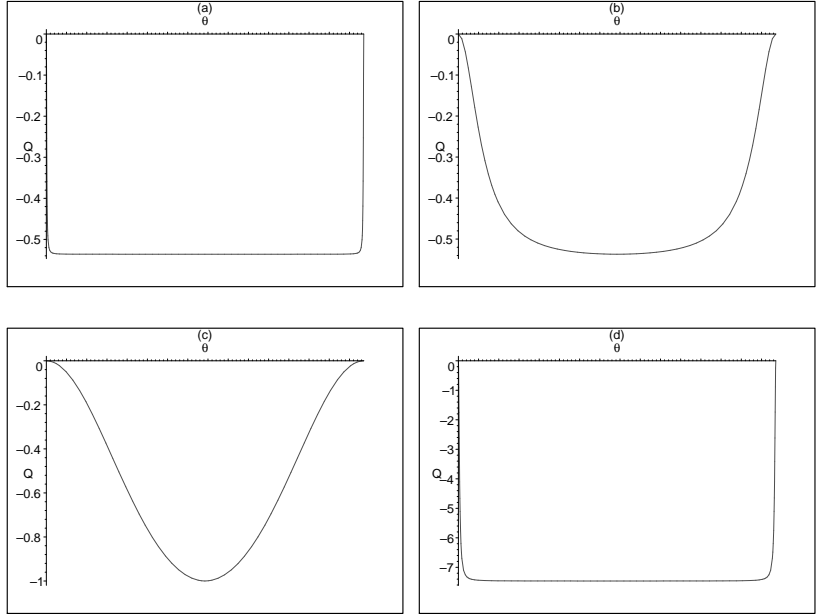

\[
\begin{array}{cc}
{\includegraphics[width = 1.50in, angle = 270]{gow24f.eps}} &
{\includegraphics[width = 1.50in, angle = 270]{gow23f.eps}}\\
{\includegraphics[width = 1.50in, angle = 270]{gow22f.eps}} &
{\includegraphics[width = 1.50in, angle = 270]{gow21f.eps}} \cr
\end{array}
\]
\caption[]{The evolution of $Q$ as a function of $\theta$ in $\tau$.
Fig. (a) corresponds to $\tau = -1.1439$, (b) to $\tau = -1.1$, (c) to
$\tau = -\ln (\pi/2)$, (d) to $\tau = +5$} 
\label{kgow2}
\end{figure}

As an example, Figs. \ref{kgow1} and \ref{kgow2} give $P - h$ ($h =
-\ln [\alpha\sqrt{1 -
\beta^2} \sin (e^{-\tau})]$)
and $Q$ as functions of $\theta$ for various values of $\tau$ for $\alpha
= 1$, $\beta = 1/2$.
The evolution of these two functions is that of
a ``spike'' that at the limiting values of $\tau$ is practically flat (except
at $\theta = 0, \pi$, where it drops off drastically), and which becomes
sharper for intermediate values of $\tau$.
It should be mentioned that the Kerr metric is a special case of what is
usually called the Kerr-Taub-NUT metric, which is geven in Ref. \cite{kramet}.
This metric has three parameters, $a$, $m$, $l$, the Kerr parameter, the mass,
and the NUT parameter respectively.
For $a = 0$ we have the metric (\ref{nut}), and for $l = 0$ we have the Kerr
metric.  This means that the Gowdy model we have given is also a special
case of the ``Taub'' region of Kerr-Taub-NUT.  However, the Taub region of
this manifold, in spite of its name, never seems to have been considered as
a cosmological model.

As we have stated several times, the models studied so far are vacuum models.
While matter-filled models may, in principle, be generated by horizon methods,
in at least some cases the models with matter have very different behavior
from the vacuum case, and may preclude naive generalizations of the method
to the non-vacuum case.  For instance, it is known that the Taub model
with fluid matter has true curvature singularities where the vacuum
case has null surfaces \cite{rysh}.  The same is true for the Kantowski-Sachs
model \cite{kon}. However, the Brill model \cite{bril} is a Taub-NUT
electrovac solution which is non-singular in the same way as the
Taub-NUT vacuum case.  In fact, the Reissner-Nordstr\o m solution with
electric charge large enough has two horizons, and between the outer and
inner horizons there is a region that can be interpreted as an electrovac
cosmology that has no curvature or electromagnetic field singularity at
the horizons.

Besides the solutions we have mentioned, there are numerous metrics that
represent gravitational fields outside some material body which in vacuum
have ``black hole'' sectors outside of some horizon.  Of course, no hair
theorems \cite{nohar} say that there are no
true vacuum black holes that are not Kerr or
Schwarzschild.  Solutions of the type mentioned above seem always to have
curvature singularities on or outside their horizons where they can be seen
by observers at infinity.  While this fact makes them poor candidates for
black holes, it makes them interesting as cosmological models inside
their horizons. The ``singularities'' of the models with curvature
singularities on their horizons will be inhomogeneous,
with part of the singularities being simple horizons, but other regions will
be true curvature singularities.  These structures might be of great
interest in the study of cosmological singularities.  Perhaps the simplest
of the metrics of this type might be the Tomimatsu-Sato
\cite{tomsat1,tomsat2} models which do have inhomogenous horizons,
but there are other known solutions, for example, one with many multipoles
and large curvature singularity regions on the horizon \cite{hern}.

\subsection{Causal Structure Methods}
This method seems to have been little used, and perhaps the most telling
reason for this is that there are very few examples where it works.  The
only example that has wide currency is that of certain Gowdy models.  If one
begins with the Einstein-Rosen waves \cite{erw}, which have the metric
\begin{equation}
ds^2 = e^{2\gamma - 2\psi}(dr^2 - dt^2) + r^2 e^{-2\psi}d\phi^2 +
e^{2\psi}dz^2,
\label{eros}
\end{equation}
with $\gamma$ and $\psi$ functions of $r$ and $t$, the Einstein equations
for this metric are
\begin{equation}
\psi_{,rr} + {{1}\over {r}}\psi_{,r} - \psi_{,tt} = 0,
\label{era}
\end{equation}
\begin{equation}
\gamma_{,r} = r(\psi_{,r}^2 + \psi_{,t}^2), \qquad \gamma_{,t} = 2r\psi_{,r}
\psi_{,t},
\label{erb}
\end{equation}
where it is well known that the equations for $\gamma$ can be integrated
directly once $\psi$ is known, since the integrability condition for
these two first-order partial differential equations is just (\ref
{era}).

If we now simply make the coordinate change $t \leftrightarrow r$, we
have
\begin{equation}
ds^2 = e^{2\gamma - 2\psi}(dt^2 - dr^2) + t^2e^{-2\psi}d\phi^2 + e^{2\psi}
dz^2.
\label{newgow}
\end{equation}
Here the only obstacle to interpreting this metric as a cosmological model
is the wrong signs in the $dr^2$ and $dt^2$ terms if we want to interpret
$\partial /\partial r$ and $\partial /\partial t$ as spacelike and timelike
vectors respectively.  If we add a complex constant to $\gamma$, $\gamma
= \tilde \gamma + i\pi$, we change the signs of $g_{tt}$ and $g_{rr}$, and
since we have done nothing more than change the names of $r$ and $t$, the
Einstein equations are unchanged (since they depend only on derivatives
of $\gamma$), and we have
\begin{equation}
-\psi_{,tt} - {{1}\over {t}}\psi_{,t} + \psi_{,rr} = 0,
\label{gowa}
\end{equation}
\begin{equation}
\tilde \gamma_{,t} = t(\psi_{,t}^2 + \psi_{,r}^2), \qquad \tilde
\gamma_{,r} = 2t\psi_{,t}\psi_{,r}
\label{gowb}
\end{equation}
These are the original equations for the unpolarized Gowdy $T^3$
cosmological model \cite{gow}.

Notice that in order to change the sign of the $dt^2 - dr^2$ term, it was
necessary to add $-i\pi$ to $\gamma$, and if the Einstein equations were
to depend on $\gamma$ in any other way than through derivatives or
$e^{\gamma}$, they would become complex and then would not necessarily have
real solutions for $\gamma$ and $\psi$.  This difficulty is a paradigm for the
problems one would encounter in trying to use a $t \leftrightarrow r$ type
of coordinate change.  One would not expect this technique to be successful
in the great majority of cases, especially if there were matter present.
This seems to be reflected in the fact that there exist few examples of this
method.  Note that the examples we have given, the plane wave of the
Introduction and the $T^3$ Gowdy model, are vacuum spacetimes, where the
temporal variable and one of the space variables appear in the form of
$-dt^2 + dz^2$ (in the Gowdy model, multiplied by a conformal factor).
While we will not try to explore this kind of idea further here,
perhaps theorems about the existence of solutions found by means
of the method could be based on this fact.

\subsection{Topological Questions}
An interesting feature of cosmological models generated by means of horizon
and causal structure methods is that for philosophical reasons one often
wants these models to have compact $t =$ const. sections, that is,
they should be closed
cosmological models.  This can be achieved for many of the solutions given
above by simply specifying the global topology without changing the local
geometry.  This has caused some problems with these models, especially
the vacuum models generated by horizon methods,
since they represent incomplete manifolds,
and when one passes through the horizon, the spacelike direction in which
the manifold would have to be closed becomes timelike, leading to the
possibility of closed timelike lines in the ``black hole'' sector.

\begin{figure}
\begin{center}
\includegraphics[width=.7\textwidth]{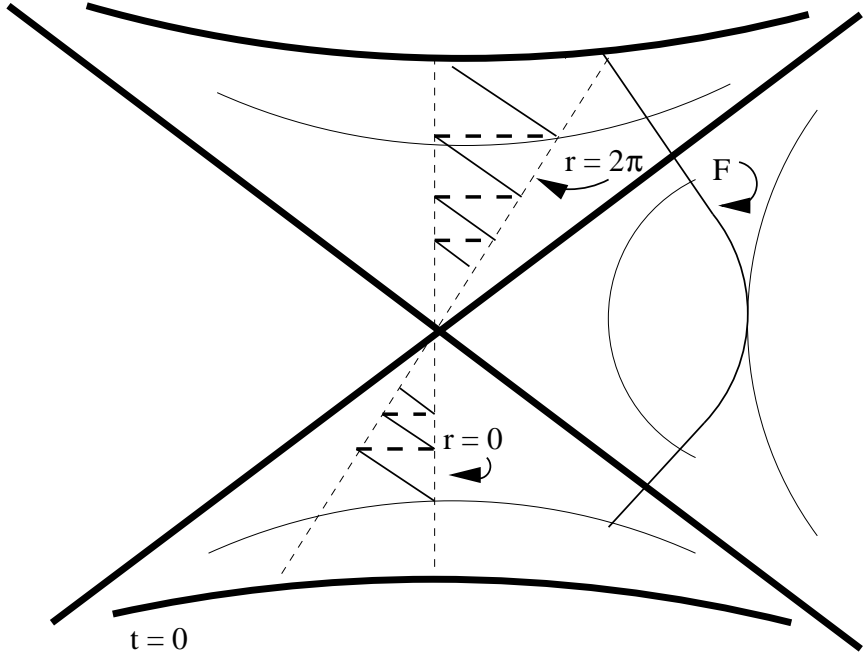}
\end{center}
\caption [] {A Kruskal diagram for geodesics
in Kantowski-Sachs-Schwarzschild.  For an open
topology the timelike geodesic $F$ can leave the cosmological region,
enter the Schwarzschild region and return to the cosmological region.
This behavior is similar to that of some geodesics in Taub-NUT.  For
the closed topology, however, geodesics wrap around between $r = 0$
and $r = 2\pi$ ($r$ being the new ``radial'' direction inside the horizon),
following the solid-dashed sawtooth path (here, as an example, for a
lightlike geodesic), which always stays inside the horizon except at the
crossing point of the past and future horizons}
\label{kdia}
\end{figure}

Perhaps the simplest model where this occurs is the Kantowski-Sachs-
Schwarzschild manifold.  If we use (\ref{ks}) for the Kantowski-Sachs model,
notice that the $\theta \varphi$ sector has a natural two-sphere topology,
and there is no obstruction to compactifying in the $r$ direction by
simply assuming that $0 \leq r \leq 2\pi$, with $r = 0$ and $r = 2\pi$
identified, giving the manifold an $S^1 \times S^2$ topology.  Of course,
if we consider the other side of the horizon, the $t \leftrightarrow r$
reparametrization means that the $r$-direction becomes the $t$-direction,
and we have $0 \leq t \leq 2\pi$ and the possibility of closed timelike
lines.  In principle one could have timelike lines which pass from the
cosmological region to the black hole region and remain trapped in that
region in an eternal closed timelike curve.  In the case of the Kantowski-
Sachs-Schwarzschild manifold it seems to be impossible for this to happen.
From Fig. \ref{kdia} one can see that timelike
geodesics which begin inside the
horizon will always stay within the horizon, simply passing through the
crossing point of the past and future horizons (a focussing point for all 
geodesics) \cite{ted}. The choice of topology
of this metric has a long history.  The original studies of this metric
assumed that $r$ ran from $-\infty$ to $+\infty$ and that the topology was
$R^1 \times S^2$.  This topology was usually used until Laflamme and
Shellard introduced the $S^1 \times S^2$ topology \cite{laf}.  Since that
time most authors have used it, but some still prefer the $R^1
\times S^2$ topology \cite{xanth}.

Probably the first solution where topological problems were noticed was
the Taub-NUT model, where the natural topology of t = const. slices
in the Taub model is $S^3$, which means that the same topology must apply
to the $t\theta \varphi$ sector of NUT space.  Here it is possible for
timelike geodesics to leave the Taub region and pass into the NUT region,
where they may be trapped in trajectories that return to the same spacetime
point \cite{chas,rysh,hell}.

For the other solutions generated by horizon methods mentioned so far, it
seems that we will have similar problems, but there has been no study of
the effect of topology changes on the causal properties of the solutions.
Even degenerate solutions such as the deSitter model and Minkowski space
can be given different topologies by means of this type of identifications.
Thus, for example, the Milne model, while it is locally flat, may be made
into a more acceptable cosmology by means of such a
global topology change, but
it would suffer from the same problem as the other metrics of this type in
that the Rindler space sector would then have an unacceptable topology.  The
Kerr-Gowdy solution has this same topological difficulty, since the interior
is supposed to have an $S^1 \times S^2$ topology, and outside the horizon
this implies an $S^1$ topology in the timelike direction.  The causal
properties of this solution with this topology seem never to have
been studied.

For metrics generated by causal structure methods, there is, in principle
no obtruction to topology changes, since there is no ``black hole'' sector
where new topologies can lead to causality problem.  The Gowdy $T^3$
model is a good example of this.  While we have not specified the ranges
of $r$, $\phi$ and $z$ in (\ref{newgow}), Gowdy, wishing to have a closed
topology for the $t =$ const. surfaces in his model, assumed that $0 \leq
r,\phi,z \leq 2\pi$ with points at $0$ and $2\pi$ identified, which gives
the model a three-torus topology.  Since there is no ``black hole'' sector
for the new manifold, there is no reason not to choose this topology.  The
only difference that this change introduces is in the boundary conditions
that $\psi$ and $\gamma$ must satisfy.  If $r$, $\phi$, and $z$ run from
$-\infty$ to $+\infty$ the solutions of Eq. (\ref{gowa})
are built up of eigenfunctions of continuous eigenvalues, while in the
Gowdy topology the eigenvalues are discrete.

Note that if we were to try to make the $t \leftrightarrow r$ change for
this topology, the Einstein-Rosen wave thus generated would be closed in the
time direction, but the Gowdy manifold is complete, and there is no
reason to concern ourselves with this possibility.

As we mentioned in the Introduction, the somewhat artificial model we gave
there may be closed in the space directions without changing the local
metric.

\section{Mapping Methods}
As we mentioned in the Introduction, the solution generating
tecniques developed in the last few decades have not often been
used in the context of cosmological models. However, it is known that
these techniques can be applied when the spacetime possesses at least two
commuting Killing vector fields.  This is exactly the case of the
cosmological models under investigation in this paper.

In the original works on solution generating techniques the central idea
was to project the spacetime on the two-dimensional hypersurface defined
by the Killing vector fields.  This projection allows one to reformulate
the spacetime metric and the corresponding field equations in such a way
that the symmetries of the differential equations can be investigated in
a straightforward manner.  An alternative but related procedure is to
derive the Ernst representation of the field equations where it is
possible to apply (with some modifications) known techniques.  In the
present work we will use this second alternative.

\subsection{Ernst Representation of the $T^3$ Gowdy Models}

In this section we will begin by considering the unpolarized $T^3$ Gowdy
model.  We will use a slightly different parametrization of these
models from that used in (\ref{newgow}) in order to make comparisons with
Refs. \cite{num}.  The unpolarized model has the metric
\begin{equation}
ds^2= e^{-\lambda/2} e^{\tau/2} ( - e^{-2\tau} d\tau^2 + d\theta^2)
+ e^{-\tau} [e^P (d\sigma + Q d\delta)^2 
+ e^{-P} d\delta^2] \ ,
\label{t3}
\end{equation}
where the functions $\lambda$,  $P$ and $Q$ depend on the coordinates 
$\tau$ and $\theta$ only, and $0 \leq \sigma, \delta, \theta \leq
 2\pi$. These spacetimes are characterized by the existence of two commuting
Killing vector fields
$\eta_1 = \partial/\partial\sigma$ and 
$\eta_2 = \partial/\partial\delta$. In the special case $Q=0$, 
the fields $\eta_1$ and $\eta_2$ become hypersurface orthogonal to
each other and the metric (\ref{t3}) describes the polarized
$T^3$ Gowdy models. 

Einstein's vacuum field equations for the $T^3$ models consist
of a set of two second order differential equations for $P$ and
$Q$
\begin{equation}
P_{,\tau\tau} - e^{-2\tau} P_{,\theta\theta} - e^{2P}(Q_{,\tau}^2 -
e^{-2\tau}Q_{,\theta}^2) = 0 \ ,
\label{t3eqp}
\end{equation}

\begin{equation}
Q_{,\tau\tau} - e^{-2\tau} Q_{,\theta\theta} + 2(P_{,\tau} Q_{,\tau} -
e^{-2\tau}P_{,\theta} Q_{,\theta}) = 0 \ ,
\label{t3eqq}
\end{equation}
and two first order differential equations for $\lambda$, 
\begin{equation}
\lambda_{,\tau} = P_{,\tau}^2 + e^{-2\tau} + e^{2P} (Q_{,\tau}^2 +
e^{-2\tau}Q_{,\theta}^2) \ , 
\label{t3eqlam1}
\end{equation}

\begin{equation}
\lambda_{,\theta} = 2(P_{,\theta} P_{,\tau} +
e^{2\tau}Q_{,\theta} Q_{,\tau}) \ .
\label{t3eqlam2}
\end{equation}
The set of equations for $\lambda$ is the equivalent of Eqs. (\ref{gowb})
and can be solved by quadratures once $P$ and $Q$ are known. 

As we mentioned, the special case of the polarized
$T^3$ model is obtained from
the metric (\ref{t3}) just by taking $Q=0$. The resulting 
field equations are easier to handle, and a general solution
for the main function $P$ can be found by separation of 
variables. In fact, let us consider $P(\tau,\theta)= T(\tau) \Theta(\theta)$; 
then, Eq. (\ref{t3eqp}) separates into 
\begin{equation}
{1 \over \Theta}{d^2\Theta \over d\theta^2} = -n^2 \ , \qquad
{\rm and} \qquad {d^2 T\over d\tau^2} + n^2 e^{-2\tau} T = 0 \ ,
\label{eqp}
\end{equation}
where $n$ is the separation constant which in this case has
to be an integer in order for the condition 
$\Theta(\theta + 2\pi) = \Theta(\theta)$ to be satisfied. 
With this assumption, the general solution for $P$ can be 
written as an infinite series of the form
\begin{equation}
P = \sum_{n=0}^\infty [ A_n \cos(n\theta) + B_n \sin(n\theta)]
[C_n J_0 (n e^{-\tau}) + D_n N_0(n e^{-\tau}) ] \ , 
\label{solpt3}
\end{equation}
where $A_n, \ B_n, \ C_n$ and $D_n$ are arbitrary constants. 
If we want to avoid singularities at $\tau = +\infty$, the 
constant $D_n$ has to vanish.  

The models described by the general $(Q\neq 0)$ 
metric (\ref{t3}) have been used
extensively for numerical investigations in classical as well
as in minisuperspace quantum gravity \cite{num}. One of the 
reasons why these investigations have used
numerical methods is because it is usually believed that 
the set of main field equations (\ref{t3eqp}) and (\ref{t3eqq}) 
is such a complicated system that analytic solutions would be
difficult to find. We will show here that it is possible to generate 
unpolarized solutions ($Q\neq 0)$ from a given polarized 
solution $(Q =0)$ by using modern solution generating 
techniques.  Of course, the new solutions will be particular solutions, and
the numerical investigations in Refs. \cite{num} were carried out with
the aim of studying the general behavior of the models near a singularity,
information that no particular solution can give.  However, families of
particular solutions can give us clues about how to set up numerical
solutions.

We can apply the solution generating techniques by writing
the field equations in such a way that the
symmetries involved can be derived and understood easily.   
To this end, we first introduce a new ``time'' coordinate 
$t=e^{-\tau}$ and a new function $R=R(t,\theta)$ 
by means of the equations
\begin{equation}
R_{,t} = t e^{2P} Q_{,\theta} \ , \qquad R_{,\theta} = t e^{2P} Q_{,t} \ .
\label{eqr}
\end{equation}

Then, the field equations (\ref{t3eqq}) and (\ref{t3eqp}) 
can be expressed as
\begin{equation}
t^2\left( P_{,tt} + {1 \over t} P_{,t} - P_{,\theta\theta} \right) 
+ e^{-2P} (R_{,t}^2 - R_{,\theta}^2) = 0 \ ,
\label{eqpr}
\end{equation}

\begin{equation}
t e^P \left( R_{,tt} + {1 \over t} R_{,t} - R_{,\theta\theta} \right)
- 2 [(t e^P)_{,t} R_{,t} - (te^P)_{,\theta} R_{,\theta} ] = 0  \ .
\label{eqqr}
\end{equation}
Furthermore, this last equation for $R$ turns out to be
identically satisfied if the integrability condition 
$R_{,t\theta} = R_{,\theta t}$ is fulfilled.

We can now introduce the complex Ernst potential $\epsilon$ and
the complex gradient operator $D$ as
\begin{equation}
\epsilon = t e^P + i R \ , \qquad {\rm and} \qquad
D= \left({\partial \over \partial t} \ , \ 
         i {\partial \over \partial \theta} \right) \ ,
\label{ernstpot}
\end{equation}
which allow us to write the main field equations in the 
{\it Ernst-like representation}
\begin{equation}
Re(\epsilon)\left(D^2\epsilon + {1\over t} D t\, D\epsilon \right)
 - (D\epsilon)^2 = 0 \ .
\label{ernstt3}
\end{equation}
It is easy to see that the field equations (\ref{eqpr}, \ref{eqqr}) 
can be obtained as the real and imaginary part of the Ernst
equation (\ref{ernstt3}), respectively. 

The importance of 
this representation is that it also can be derived from
a Lagrangian by means of a variational principle. In turn, 
that Lagrangian may be interpreted as a metric Lagrangian
defined in a two dimensional Riemannian space 
(the potential space), the coordinates of which are the  real and
imaginary parts of the Ernst potential. Applying the variational 
principle in the potential space on the metric Lagrangian, one
obtains the corresponding geodesic equations, which turn out
to be equivalent to the Ernst equation (\ref{ernstt3}). Hence,
to investigate the symmetries of the Ernst equation one can 
study infinitesimal transformations which leave invariant 
the geodesic equations in the the potential space. In particular, 
the transformations associated with the Killing vector fields 
of the metric in the potential space leave the corresponding
geodesic equations invariant. One could think of a solution
of the Ernst equation as a geodesic in the potential space, 
and the Killing vectors of the metric as transformations that
starting from a given geodesic lead to a different geodesic,
i.e., to a different solution of the Ernst equation. This 
is the basic idea behind some of the known solution generating
techniques. 

We will now derive a simple but illustrative symmetry of the 
Ernst equations which allows to generate new solutions. To 
this end, we introduce a new complex potential $\xi = 
\xi(t, \theta)$ by means of the relationship
\begin{equation}
\epsilon = { 1 - \xi \over 1 + \xi} \ .
\label{xi}
\end{equation}
Then, the Ernst equation (\ref{ernstt3}) transforms into
\begin{equation}
(1 - \xi\xi^*) \left( D^2 \xi + {1 \over t} Dt \, D\xi \right)
+ 2\xi^* (D\xi)^2 = 0 \ , 
\label{eqxi}
\end{equation}
where $\xi^*$ represents the complex conjugate potential. 
Furthermore, we introduce new coordinates $x$ and $y$ by
\begin{equation}
t^2 = c^2 (1-x^2)(1-y^2)\ , \qquad \theta = c x y \ ,
\label{tra}
\end{equation}
where $c$ is a real constant. In these coordinates, the main field 
equation (\ref{eqxi}) can be written in the following form 
\begin{equation}
(1-\xi\xi^*)\{ [(1-x^2)\xi_{,x}]_{,x} - [(1-y^2)\xi_{,y}]_{,y}\}
+2\xi^* [ (1-x^2)\xi_{,x}^2 - (1-y^2)\xi_{,y}^2] = 0 \ ,
\label{eqxixy}
\end{equation}
which explicity shows the invariance with respect to the 
change of coordinates $x\leftrightarrow y$, i.e., if 
$\xi(x,y)$ is a solution of (\ref{eqxixy}) then $\xi(y,x)$
is also a solution. The simplest solution of Eq. (\ref{eqxixy})
is $\xi^{-1} = x$, so $\xi^{-1} = y$ is also a solution. A linear
combination of these two solutions 
\begin{equation}
\xi^{-1} = a x + i b y
\label{xiks}
\end{equation}
turns out to also be a solution if the condition $a^2 + b^2 = 1$ 
is satisfied.
In this way it is relatively easy to generate new solutions. Once
the potential $\xi$ is known one can calculate the functions 
$P$ and $R$ algebraically by means of Equations (\ref{xi}) and 
(\ref{ernstpot}). Then, Eq. (\ref{eqr}) may be integrated to yield
$Q$. Finally, the solution for the function $\lambda$ may be 
calculated by quadratures. 

It is interesting to note that one can derive some of the
main properties of the Gowdy model just by inspecting the explicit
form of the Ernst potential. If the Ernst potential $\epsilon$
is real, then the function $R$ vanishes and, therefore,
the metric function $Q$ is a constant which can be absorbed by
means of a suitable coordinate transformation. Thus, a real Ernst
potential corresponds to a polarized Gowdy model.
In the case of complex Ernst potential with nontrivial
imaginary part (not proportional to its real part), it is
guaranteed that the resulting metric corresponds to an 
unpolarized Gowdy model ($Q\neq 0)$. For instance, even 
the simple solution (\ref{xiks}) will lead to 
a nontrivial unpolarized solution with $Q\neq 0$. 

\subsection{Generation of New Solutions}

In this section we will briefly describe the way new solutions
can be generated by using HKX transformations. Let us assume that
a solution for the {\it polarized} Gowdy model is given by 
$P=P_0(t,\theta)$  and $\lambda=\lambda_0(t,\theta)$. 
The corresponding Ernst potential is  real, and we denote it
 by $\epsilon_0 = t e^{P_0}$.  

A HKX transformation acting on $\epsilon_0$ 
generates a new complex Ernst potential $\epsilon$ which will
correspond to an unpolarized Gowdy model only if its imaginary
part is not proportional to its real part. This will depend
on the explicit form of the initial Ernst potential $\epsilon_0$.
However, if we apply two different HKX transformations it can
be shown that the resulting Ernst potential is nontrivial. 
For this reason, we present here the result of the action 
of two HKX transformations on the real Ernst potential 
$\epsilon_0$. For the sake of simplicity, we use the coordinates 
$x$ and $y$ defined in Eq. (\ref{tra}) in which the new Ernst 
potential can be written as \cite{que1}
\begin{equation}
\epsilon = \epsilon_0 { 
x(1-\mu_1\mu_2)+iy(\mu_1+\mu_2)+(1+ \mu_1\mu_2)+ i(\mu_1-\mu_2)
\over 
x(1-\mu_1\mu_2)+iy(\mu_1+\mu_2)+(1+\mu_1\mu_2)- i(\mu_1-\mu_2) },
\label{newernst}
\end{equation}
where we have introduced new functions $\mu_1$ and $\mu_2$ 
defined by
\begin{equation}
\mu_1 = \alpha_1 e^{2\beta_-}  \qquad {
\rm and } 
\qquad 
\mu_2 = \alpha_2 e^{2\beta_+}  \ .
\label{mus}
\end{equation}
Here $\alpha_1$ and $\alpha_2$ are real constants and $\beta_\pm$
are functions which satisfy a set of two first order differential
equations 
\begin{equation}
(x \mp y)(\beta_\pm)_{,x} = 
(1 \mp xy){\epsilon_{0,x}\over \epsilon_0}  \mp 
(1-y^2)   {\epsilon_{0,y}\over \epsilon_0} 
\label{betax} \ ,
\end{equation}

\begin{equation}
(x \mp y)(\beta_\pm)_{,y} = 
(1 \mp xy){\epsilon_{0,x}\over \epsilon_0}  \pm 
(1-x^2)   {\epsilon_{0,y}\over \epsilon_0} 
\label{betay} \ .
\end{equation}

Thus, the generation of new solutions reduces to the integration
of the differential equations for $\beta_\pm$. The new constants
$\alpha_1$ and $\alpha_2$ have been introduced by the two HKX 
transformations. For vanishing $\alpha_1$ and $\alpha_2$, the 
potential (\ref{newernst}) reduces to the original Ernst potential
of the polarized Gowdy model. 

Equations (\ref{betax}) and (\ref{betay}) allow us to
generate new {\it unpolarized} solutions starting from a given
{\it polarized} Gowdy model. The derivation of explicit solutions
implies the integration of a set of two first order differential
equations for a function which determines the Ernst potential of the new
solution. In turn, from the Ernst potential one can obtain
the new metric functions which completely determine the
spacetime of the new cosmological model.  In order to complete
this procedure it is necessary to carry out lenghtly but
straightforward calculations. We are attempting to calculate
the unpolarized solution corresponding to the general solution
(\ref{solpt3}) of the polarized $T^3$ Gowdy model. 

\section{Conclusions and Suggestions for Further Research}
We have given several examples of the three methods listed in the
Introduction.  It is obvious that these three approaches are far from
exhausted, and may give a number of interesting exact solutions to
known types of cosmological models and suggest new types of models 
of interest.  It is probable that the horizon and mapping methods
may be more productive than causal structure methods.

It seems that horizon methods have barely touched the surface of
possible solutions.  There are a large number of exact solutions of
``black hole'' type given in Ref. \cite{kramet} that are candidates
for generating cosmological models inside their horizons.  We have looked 
at the Kantowski-Sachs-Schwarzschild, Taub-NUT and Kerr manifolds in
some detail.  We have also mentioned as possibilities the Tomimatsu-Sato
and Quevedo class of solutions which should lead to models with mixed
singularities, part null surface and part curvature singularity.  There
are large classes of vacuum type D solutions that might give interesting
cosmologies \cite{kramet}.  Even such well known
solutions as Kerr-Taub-NUT do not seem to have been investigated
thoroughly as cosmological models.

Models with matter seem hardly to have been touched.  While fluid models
may be difficult to generate, the example of electrovac universes shows
us that there are many possible cosmologies of this type.  Even the
interior of the Reissner-Nordstr\o m solution does not seem to have 
been investigated in this context. It would also be interesting to
investigate the possibility of obtaining useful cosmological models
from the most general type D electrovac solutions with cosmological
constant, found by Debever et al. \cite{deb}, which contains 13
different parameters and includes the Kerr-Newman solution as a 
special case.  
Even the Kerr-Newman solution \cite{newman2} would generate an $R^1 \times
S^2$ (and even, perhaps, an $S^1 \times S^2$) Gowdy solution, equivalent
to, or a generalization of, the solutions of Carmeli, Charach and
Malin \cite{cmal} for $T^3$ models.  Other types of solutions, such as
scalar field models do not seem to have been investigated at all in this
context.

If anything, mapping methods seem even more underutilized in cosmology.
The example we have seen is one of many that come to mind.  For the
Gowdy models one could use these methods to investigate the $S^1 \times
S^2$ models and perhaps generate unkown solutions.  The corresponding
Ernst potential of the field equations present some technical difficulties
related to the specific topology of the model.  This problem is currently
under investigation.  For this case, it would also be interesting to
generate new unpolarized solutions starting from the general polarized
solution, which in the case of $S^1 \times S^2$ models, just as in the
$T^3$ models, can also be represented as an infinite series of
eigenfunctions since the corresponding field equations reduce to 
separable linear second order differential equations.  

Solutions with
with true curvature singularities on possible inner and outer horizons
might be especially interesting.  An example of this idea was given by
Moncrief \cite{mon}, who used the Geroch group to generate a solution
that had a curvature singularity in place of a horizon in the 
Kerr-Taub-NUT case. 

There will certainly be many opportunities to apply all three of the
techniques discussed in the article to the generation of new
cosmologies.

\section*{Acknowledgments}
We wish to thank T. Jacobson, V. Moncrief and O. Obregon for useful
discussions.  This work was supported in part by CONACyT grant
3567E, and DGAPA-UNAM grants 121298 and IN106097.

\end{document}